\documentstyle[12pt]{article}
\newcommand{\BE}{\begin{equation}}
\newcommand{\EE}{\end{equation}}
\newcommand{\BA}{\begin{eqnarray}}
\newcommand{\EA}{\end{eqnarray}}
\newcommand{\doublespace}{\renewcommand{\baselinestretch}{1.5}\large\normalsize}
\doublespace
\begin{document}


\begin{center}

{\bf Acoustic scattering by periodic arrays of air-bubbles}\\ \ \\

Zhen Ye

\end{center}

\begin{center}
\verb+Author+: Zhen Ye\\ \verb+Professional address+: Wave
Phenomena Laboratory, Department of Physics, National Central
University, Chungli, Taiwan, Republic of China\\ \verb+Email+:
zhen@phy.ncu.edu.tw\\ \verb+Fax+: +886-3-4251175; \verb+Phone+:
+886-930012632\\ \verb+Running title+: Acoustic scattering by
bubble arrays
\end{center}

\newpage

\begin{abstract}

This paper considers acoustic scattering by and propagation
through line and plane arrays of air-bubbles in liquid media. The
self-consistent method is used to derive the effective scattering
function of a single bubble embedded in the arrays, incorporating
all multiple scattering processes. For the line case, an exact
result is derived. In the plane array situation, only an
approximate analytic result is possible. Numerical computations
have been carried out to show the multiple scattering effects on
wave scattering.  It is shown that depending on the distance
between bubbles the resonance peak of a single bubble can either
be broadened or narrowed due to multiple scattering and it shows
an oscillatory behavior as the distance changes.  Meanwhile, the
peak scattering amplitude is also be either enhanced or reduced.
The previously predicted strong enhancement, however, is not
evident. For plane arrays, the usual resonant scattering of a
single bubble in absence of other bubbles can be suppressed by
multiple scattering when the distance between bubbles is
sufficiently small. As the distance increases, the resonant
scattering starts to appear, and the resonance peak position is
alternately shifted towards higher and lower values. Moreover, it
is predicted that wave propagation through a plane bubble array
can be significantly inhibited in a range of frequencies slightly
higher than the natural frequency of a single bubble, possibly a
useful feature for noise screening. The ambiguities in the
previous results are pointed out.

\end{abstract}

\newpage

\section{Introduction}

When propagating through media containing many scatterers, waves
will be scattered by each scatterer.  The scattered wave will be
again scattered by other scatterers.  Such a process will be
repeated to establish an infinite recursive pattern of
rescattering between scatterers, forming a course of multiple
scattering. Because of multiple scattering, the overall scattering
effect in the system may not be represented simply by the sum of
the effects of individual scatterers in isolation. It has now
become well-known that multiple scattering of waves is responsible
for a wide range of fascinating phenomena. This includes, on large
scales, twinkling light in the evening sky, modulation of ambient
sound at ocean surfaces \cite{vagle}, and acoustic scintillation
from turbulent flows\cite{2} and fish schools\cite{3}.  On smaller
scales, phenomena such as white paint, random laser\cite{4},
electrical resistivity, photonic\cite{5} and sonic
bandgaps\cite{Ye} in periodic structures also find their roots in
wave multiple scattering. Even more interesting, perhaps, multiple
scattering may lead to a phase transition in wave propagation,
that is, due to multiple scattering propagating waves may be
trapped in space and will remain confined in the neighborhood of
initial site until dissipated.  In the meantime, individual
scatterers shows an amazing collective behavior; such a collective
behavior effectively prevent waves from propagation and yields the
phenomenon of Anderson wave localization\cite{6,6a}.

Tremendous effort has been devoted to the study of multiple
scattering of waves and a large body of literature exists (Refer
to, for example, the monograph by Ishimaru\cite{8}). The work of
Poldy\cite{Foldy}, Lax\cite{Lax}, Waterman et al.\cite{Waterman},
Twersky\cite{Twersky}, and many others serves as a cornerstone to
the subject and provides various schemes describing multiple
scattering processes in a number of situations of interest.  In a
series of articles, Foldy and Twersky described the multiple
scattering of waves in media containing arbitrary scatterers by a
set of self-consistent equation.  If not impossible, the exact
solution to such a set of coupled equations is difficult to
obtain. Certain approximations, such as the perturbation series in
the diagrammatic method\cite{Ding}, generally have to be resorted
to.

An exact description of multiple scattering is only possible in
several simple systems. Multiple scattering of acoustic waves by a
finite number of air-filled bubbles in liquids has been one of
such rare systems and poses a useful model system to study wave
multiple scattering\cite{Zabolotskaya}.  A rigorous treatment of
multiple scattering not only provides definite but also new
insight into phenomena associated with multiple scattering. For
example, the recent numerical investigation\cite{6,6a,AAD} has
shed further light on the aforementioned phenomenon of wave
localization which could not be possibly obtained within the realm
of approximations. Despite the success, however, the research is
purely numerical and has been limited to the case of a finite
number of air-bubbles. It would be desirable to explore more
complicated and practical situations involving an infinite number
of scatterers and pursue the conditions in which an analytic
description of multiple scattering can be obtained, thus rendering
a justification of numerical results extended for infinite
scatterers.  This article presents one of such attempts.

The problem considered here is multiple scattering of acoustic
waves by regular arrays of air-bubbles in a liquid. Two situations
will be deliberated: linear and planar arrays.  It is shown that
the two cases have the closed form solutions.  These two simple
situations are chosen so as to display the physical essence in a
most explicit form. There are a number of earlier works on these
circumstances. Weston\cite{Weston} first considered the frequency
response of air-bubbles forming linear and planar arrays.  Weston
derived approximate formulas for sound scattering by an air-bubble
embedded in the arrays and predicted that the line array of
air-bubbles behaves like a cylindrical bubble: the sharp sphere
resonance, a well-known feature for a single spherical
bubble\cite{Clay}, is suppressed and a broader resonance at a
lower frequency appears. He further showed that a plane array of
air-bubbles behaves like a plane screen of gas - there is no
resonance at all. Later, bubbles in linear arrays were further
numerically studied in \cite{Feuillade} for the case that the
incident wave is perpendicular to the linear axis.  In contrast to
Weston, his results indicate that depending on the spacing between
bubbles, a line of bubbles need not necessarily lead to an
increase in damping for the ensemble. However, the results in
\cite{Feuillade} are not only too limited but are distracted by
errors. Tolstoy and Tolstoy\cite{18,19} also considered line and
plane arrays of air-bubbles. They predicted that pronounced
partial resonances can be observed in both systems. Their results,
however, have been questioned\cite{20,20a,20b}. In short, the
previous results are in discrepancy. A definite and careful
investigation is clearly needed.

This paper considers further the problem of line and plane arrays
of resonant monopole scatterers like the air-bubbles. The
self-consistent approach from Foldy\cite{Foldy} and
Twersky\cite{Twersky} will be followed to derive a set of coupled
equations for which exact analytic solutions are obtained.  It
will be shown that the total acoustic scattering by arrays of
bubbles can be expressed in terms of the scattering from
individual bubbles.  The effects of multiple scattering is
represented by an effective scattering function of a single bubble
embedded in the array.  Both line and plane line arrays will be
considered. Wherever appropriate, comparison with the previous
results will be made. We note that the present investigation is
limited to the linear response of air-bubbles. When the
stimulation field is too strong, therefore the interaction between
bubbles can be very large, the linear response approximation may
fail.

\section{Theory}

The problem considered here is illustrated by Figs.~1 and 2. In
Fig.~1, a unit plane wave $e^{i\vec{k}\cdot\vec{r}}$ is incident
on a line array of identical air-bubbles.  The incident wave makes
an angle of $\theta$ with the line of the array. The distance
between two neighboring bubbles is $d$. The bubble radius is taken
as $a$. Fig.~2 shows that the identical bubbles form a square
lattice in the $x - y$ plane with lattice constant $d$. The
incident wave is in the direction denoted by $\theta$ and $\phi$
in the spherical coordinates.

In absence of other bubbles, the scattered wave from a single
bubble can be written as \BE
p^i_s=p_0(\vec{r}_i)f\frac{e^{ik|\vec{r}-\vec{r}_i|}}{|\vec{r}-\vec{r}_i|},
\EE in which $p_0(\vec{r}_i)$ is the incident wave at the bubble
located at $\vec{r}_i$ and f is the scattering function of the
single bubble. It has been found numerically\cite{Ye1997} that
when $ka < 0.35$, to which the following discussion is restricted,
the scattering function is isotropic and given by \BE
f=\frac{a}{\frac{\omega_0^2}{\omega^2}-1 -i\delta}, \EE where
$\omega_0$ is the natural frequency of the bubble and $\delta$ is
the damping factor of the bubble including radiation, thermal
exchange, and viscosity effects[Refer to Appendix 6 in
Ref.~\cite{Clay}].

When many bubbles are present, the scattered wave from the $i$-th
bubble is a linear response to the total incident wave and the
scattered wave from other bubbles, and therefore can be written as
\BE p_s^i(\vec{r}) = f\left(p_0(\vec{r}_i) + \sum_{j\neq i}^\infty
p_s^j(\vec{r_i})\right)\frac{e^{ik|\vec{r}-\vec{r}_i|}}{|\vec{r}-\vec{r}_i|}.
\label{eq:sf} \EE We define an effective scattering function for
each bubble as \BE p_s^i(\vec{r})=p_0(\vec{r}_i)F_i
\frac{e^{ik|\vec{r}-\vec{r}_i|}}{|\vec{r}-\vec{r}_i|}.
\label{eq:effective}\EE Due to the symmetry, all bubbles have the
same effective scattering function, i.~e. $F_1 = F_2 = \cdots =
F_i = \cdots = F$.

Substituting Eq.~(\ref{eq:effective}) into Eq.~(\ref{eq:sf}), we
obtain \BE F = \frac{1}{f^{-1} - \sum_{j\neq i}^\infty
\frac{e^{ik|\vec{r}-\vec{r}_i|}}{|\vec{r}-\vec{r}_i|}e^{ik|\vec{r}-\vec{r}_i|}}.
\label{eq:F} \EE The total scattered wave will be \BE p_s(\vec{r})
= \sum_{i=1}^N p_s^i(\vec{r}) = F \sum_{i=1}^\infty p_0(\vec{r}_i)
\frac{e^{ik|\vec{r}-\vec{r}_i|}}{|\vec{r}-\vec{r}_i|}.
\label{eq:ps}\EE

\subsection{Infinite line arrays}

{\it Effective scattering function}. In the line array case, we
have from Eq.~(\ref{eq:F}) \BE F = \frac{1}{f^{-1} -
2k\sum_{n=1}^\infty \frac{e^{ikdn}}{nkd}\cos(kdn\cos\theta)},
\label{eq:F2} \EE where $n$ takes positive integers. Equation
(\ref{eq:F2}) is equivalent to the result previously encountered
by Weston\cite{Weston}. Define \BE I \equiv 2\sum_{n=1}^\infty
\frac{e^{ikdn}}{nkd}\cos(kdn\cos\theta). \EE This can be evaluated
as \BA I &=&
\frac{1}{kd}\{-\ln(1-\cos(1-\cos[kd(1+\cos\theta)])(1-\cos[kd(1-\cos\theta)])
\nonumber \\ & & +
i(2\pi-[kd(1+\cos\theta)]-[kd(1-\cos\theta)])\}, \label{eq:I} \EA
where $[x]$ means $2\pi$ modulo of $x$; therefore
$[kd(1-\cos\theta)]+[kd(1+\cos\theta)]\neq 2kd.$ Writing $x=2n\pi
+ x'$ with $n$ being an integer and $x'$ limited to $(0, 2\pi)$,
then $[x]=x'$. With (\ref{eq:I}), Eq.~(\ref{eq:F2}) represents the
exact solution. The result in Eq.~(\ref{eq:I}) differs from the
previously published result \cite{18} by a factor of $\sqrt{2}$ in
the logarithm and by the modulo values.

{\it Total scattered field.} The total scattered wave can be
evaluated in the far field limit, $r >> d$ with $r$ being the
perpendicular distance from the field point to the line.  In this
limit, the summation in Eq.~(\ref{eq:ps}) can be converted into an
integral. The resulting formula is \BE p_s(r) =
i\pi\frac{F}{d}H_0^{(1)}(kr\sin\theta), \label{eq:ps2}\EE where
$H_0^{(1)}(x)$ is the zero-th order type one Hankel function.

With Eq.~(\ref{eq:F2}), we see from Eq.~(\ref{eq:ps2}) that the
scattered wave is isotropic in the plane perpendicular to the line
and it depends on the incident angle and on the perpendicular
distance. Eq.~(\ref{eq:ps2}) also indicates that the scattering by
the array is characterized by the effective scattering function of
each individual bubble. This may partially resolve the debate
between Tolstoy et al. and Twersky\cite{18,20,20a,20b}. Tolstoy et
al. used the effective scattering function to study the
superresonance behavior of a bubble array. Twersky alternatively
defined a scattering amplitude of the whole array and claimed that
this is only observable.  He further stated that an individual
scattering amplitude is not observable and numerical computations
for an individual scattering function do not represent physically
observable data. Tolstoy et al. argued that the characteristics of
a single bubble can be inferred at near field. From the present
approach, we see that in the case of an infinite line array, the
individual effective scattering function does provide useful
information and represent observable data at far field as well. In
fact, Eq.~(\ref{eq:ps2}) will be valid as long as $r >> d$ and $Nd
>> r$ with $N$ being the total number of the scatterers. Therefore
even for a finite number of scatterers, the individual scattering
function may still delineate the observables.

{\it Modified natural frequency and damping rate}. With
Eq.~(\ref{eq:I}), the scattering function F is solved as \BE
F=\frac{1}{\frac{\omega^2_0}{\omega^2}-1 - i\delta - kaI}. \EE
Thus the new resonance peak and the damping rate $(\delta_R)$ are
determined from \BE \frac{\omega_0^2}{\omega^2}-1 -kaI_R = 0, \ \
\ \delta_R = \delta+kaI_M,\EE respectively. Here $I_R$ and $I_M$
represent the real and the imaginary parts of $I$ separately: \BE
I_R =
\frac{-1}{2kd}\ln(1-\cos[kd(1+\cos\theta)])(1-\cos[kd(1-\cos\theta)]),
\EE and \BE I_M=\frac{1}{2kd}(2\pi-[kd(1+\cos\theta)] -
[kd(1-\cos\theta)]). \EE

\subsection{Plane arrays}

{\it Effective scattering function}.  In this case, the origin can
be fixed at the position of one of the bubbles.  The effective
scattering function of a single bubble will be \BE F =
\frac{1}{f^{-1}- \sum_j'
\frac{e^{ik|\vec{r}_j|}}{|\vec{r}_j|}e^{i\vec{k}\cdot\vec{r}_j}},
\label{eq:15} \EE where the summation runs over all the bubbles
except the one at the origin. Unfortunately, the summation in
Eq.~(\ref{eq:15}) cannot yet be made into a simple closed form and
in general has to be evaluated numerically. For simplicity, we
consider the case $\phi=0$ [Fig. 2], i. e. the incidence is in the
$x - z$ plane. Define a quantity I: \BE I \equiv \sum_j'
\frac{e^{ik|\vec{r}_j|}}{|\vec{r}_j|}e^{i\vec{k}\cdot\vec{r}_j}.
\EE The summation can be approximated as \BE I \approx
\frac{1}{kd^2}\int_d^\infty d^2r
\frac{e^{ikr}}{r}e^{i\vec{k}\cdot\vec{r}}=\frac{2\pi}{(kd)^2}\left(\frac{i}{\sin\theta}
-
\int_0^{kd} dx e^{ix}J_0(x\cos\theta)\right), \label{eq:17}\EE
where $J_0$ is the zero-th order Bessel function of the first
kind. For normal incidence, $\theta = \pi/2$, we get approximately
\BE I = \frac{2\pi i}{(kd)^2}e^{ikd}.\EE The effective scattering
function is thus \BE F= \frac{1}{f^{-1} - kI} = \frac{1}{f^{-1} -
\frac{2\pi k i}{(kd)^2}e^{ikd}}. \label{eq:19} \EE As will be
shown later, the approximated result in Eq.~(\ref{eq:19}) is
reasonable.

{\it Natural frequency and damping rate}.  Similar to the line
case, the resonance frequency incorporating multiple scattering is
determined by \BE \frac{\omega_0^2}{\omega^2} -1 -kaI_R = 0, \EE
which is reduced to \BE \frac{\omega_0^2}{\omega^2} -1 +
\frac{2\pi ka}{(kd)^2}\sin(kd) = 0, \EE for the normal incidence.
The corresponding damping rate is \BE \delta_R = \delta +
\frac{2\pi ka}{(kd)^2}\cos(kd).\EE

{\it Transmission through bubble screens}.  Consider the normal
incidence. The transmitted wave through one plane array of bubbles
\BE p_F = e^{i\vec{k}\cdot\vec{r}} + \sum_{i=1}^N
F\frac{e^{ik|\vec{r}-\vec{r}_i|}}{|\vec{r}-\vec{r}_i|}. \EE Assume
that the incident wave is along the $z$-axis and we use the
cylindrical coordinates; here the incident direction is the axis
of symmetry. Considering the far field limit as in the line case,
we obtain \BE p_F = \left(1+\frac{2\pi ik
F}{(kd)^2})\right)e^{ikz}. \label{eq:fw}\EE The backscattered wave
is \BE p_B = \frac{2\pi ik F}{(kd)^2}e^{-ikz}. \EE In deriving
Eq.~(\ref{eq:fw}), we used the following result\BE \int_0^\infty
dt \frac{t e^{it_0\sqrt{1+t^2}}}{\sqrt{1+t^2}} =
\frac{1}{it_0}e^{it_0}, \EE which can be obtained by the method of
change of variables.

{\it Validation of the approximation and energy conservation}. As
we can see from the above, in the derivation of Eq.~(\ref{eq:17}),
we used the approximation which converts the summation into the
integral. A way to verify the approximation is to apply the energy
conservation law.  Write \BE T = \frac{2\pi ikF}{(kd)^2}.
\label{eq:27}\EE Obviously $|1+T|$ and $|T|$ represent the
transmission and reflection coefficients respectively. Note that
the present reflection coefficient differs from that of Weston
(Refer to Eq.~(29) in Ref.~\cite{Weston}. In the limit of $kd <<
1$, Weston obtained \BE |T|^2 = \frac{1}{1+
\left(\frac{\omega_0^2}{\omega^2}\frac{kd^2}{2\pi a}\right)^2},\EE
while the present approach yields \BE |T|^2 = \frac{1}{1+
\left((\frac{\omega_0^2}{\omega^2}-1)\frac{kd^2}{2\pi
a}\right)^2}.\EE 1.  There are some further approximations in
Weston's derivation. His result is more valid for frequencies
significantly below the nature frequency of the bubbles.

The energy conservation law requires, \BE |1+T|^2 + |T|^2 + A = 1,
\EE where $A$ denotes the absorption due to the thermal exchange
and viscosity: \BE A = \frac{4\pi}{kd^2}\mbox{Im}[F] -
\frac{4\pi}{d^2}|F|^2, \label{eq:A}\EE where `Im' means taking the
imaginary part. In deriving Eq.~(\ref{eq:A}), we used the optical
theorem\cite{8}. It can be shown that the approximate result from
Eq.~(\ref{eq:19}) satisfies the energy conservation law to a few
percentage, hereby providing a justification of the approximation.

\section{Numerical results}

The above theoretical results will be illustrated by numerical
examples in this section.  We consider that bubbles arrays are
placed in water. The following parameters are used in the
computation: the sound speeds of the air inside the bubble and of
the surrounding water are 340 m/s and 1500 m/s, the mass densities
for air and water are 1.29 and 1000 kg/m$^3$ respectively. The
thermal and viscosity coefficients are taken from Appendix 6 in
\cite{Clay}.

\subsection{Line arrays}

In Fig.~3, the scattering strength $|F|$ of a single bubble from
Eq.~(\ref{eq:F2}) is plotted versus $ka$ for various bubble
separations, with comparison to the scattering strength of the
single bubble in absence of other bubbles.  The incidence is
perpendicular to the line. The bubble radius is 1 mm. Figures (b),
(d), and (f) are respectively replots of (a), (c), and (e) in the
normalized scales, i.e. normalized by the maximum of the
scattering strength. From these figures, we see that (1) when the
separation between bubbles is small, the multiple scattering
heavily suppresses the scattering strength of the single bubble;
(2) the resonance peak is red shifted (shifted towards lower
values); (3) the resonance peak can either broadened or narrowed,
depending on the separation; (4) the multiple scattering effects
are negligible when the separation exceeds a certain value.

Fig.~4 plots the effective scattering strength as a function of
$ka$ for three incident angles.  Here we see that as the incidence
angle deviates from the normal direction, the resonance peak will
be further reduced and the peak position is more shifted towards
lower frequencies.  The previously predicted quasi-resonance
\cite{19} is not evident. As expected, when the separation is
sufficiently large, the results reduce to that of a single bubble
without the presence of other bubbles.

Fig.~5 plots the relative natural frequency and damping shifts
with respect to the case without multiple scattering as a function
of the bubble separation.  Two bubble sizes and two incident
angles are assumed.  The results are shown to be almost the same
for the two bubble sizes. While the frequency is always shifted to
a lower value, the damping shift shows an interesting regular
oscillatory feature. As the incidence is deviated from the normal
direction, the oscillation period is reduced. At certain ranges of
bubble separations, the damping is reduced by the multiple
scattering, indicating that the resonance peak is narrowed.  The
special cases considered in Ref.~\cite{Feuillade} fall in these
ranges.

The relative peak scattering amplitude is plotted as a function of
bubble separation in Fig.~6 for two bubble sizes and two incident
angles. Again, the oscillatory features appear, in line with that
shown in Fig.~5. For most separations, the peak amplitude is
reduced by multiple scattering.  In certain ranges of separation,
however, the peak scattering strength is enhanced for the larger
bubble case, but not as much as predicted by Tolstoy\cite{19}. For
the smaller bubbles, the peak amplitude is always reduced by
multiple scattering.  It is worth noting that the thermal and
viscosity effects, which are more important for smaller bubbles,
are not considered by Tolstoy.  This may explain partially the
discrepancy.

\subsection{Plane array}

In this situation, only normal incidence is considered.  The
effective scattering strength $|F|$ is plotted against $ka$ for
various lattice constants ($d/a$) in Fig.~7. In the plots, the
dashed lines refer to the scattering strength of a single free
bubble.  The multiple scattering effects are stronger than in the
line case.  As we can see, the resonance scattering of the single
bubble is fully suppressed when the bubbles are closed packed, in
agreement with Weston's results\cite{Weston}.  As the bubble
separation increases, however, the resonance picture starts to
appear and the multiple scattering effects decreases.

Consider the cases that the resonant scattering appears.  In
contrast to the line array case, the natural frequency is shifted
to higher values when the separation is less than certain amount.
In some ranges of separation, the resonance peak is shifted to a
lower value slightly.  This is shown by Fig.~8 for bubble radii 1
and 0.1 mm separately.  In this figure, the effects of multiple
scattering on the damping is also presented. For $d/a$ smaller
than about 100, the damping is significantly increased by multiple
scattering, implying peak broadening.  There is a range of bubble
separations, however, the damping is in fact reduced by multiple
scattering; this occurs for $d/a$ between 100 and 400.

The relative scattering amplitude at resonance with respect to
that of a single free bubble is plotted versus $d/a$ for two
bubble sizes in Fig.~9. Again, an oscillatory property surfaces.
The reduction and enhancement in scattering strength alternate, as
the bubble separation varies.  The larger the bubble, the larger
is the enhancement factor; it can be as large as 1.4 for $a = 1$
mm.

We also studied the transmission through the plane array.  The
transmission coefficient $|1 + T|$ is shown in Fig.~10 as a
function of $ka$ for four bubble separations for the with and
without thermal and viscosity dissipation. As shown in the plots,
the propagation can be significantly blocked by the plane array
for frequencies slightly above the resonance of a single free
bubble; the vertical bar indicates the resonance position of the
single free bubble. As the separation increases, the range of
frequencies in which the transmission is inhibited decreases,
until disappears.  Such an inhibition property may be useful for
utilizing bubble layers as a noise screen.  The inhibition also
play a role in defining acoustic wave localization in bubbly
liquids\cite{AAD}. When the thermal or viscosity damping is not
included, the transmission is shown to be enhanced at certain
bubble separations, as illustrated by Fig.~10(c). This is against
the energy conservation, indicating the failure of the formula
(\ref{eq:27}) in this case.

\section{Summary}

In summary, we considered acoustic scattering from regular arrays
of air-bubbles for low frequencies, i.e. $ka <<1$.  The
self-consistent method is used to derive the effective scattering
function of a single bubble embedded in the arrays, including all
multiple scattering processes. The total scattered wave is
expressed in term of this effective scattering function.  An exact
solution is presented for the case of line arrays.  For the plane
arrays, an approximate result is given. The approximation is
justified in view of energy conservation. The numerical results
show that depending on the distance between bubbles the resonance
peak of a single free bubble can either be broadened or narrowed
due to multiple scattering and shows an oscillatory feature as the
distance is increased.  In the same spirit, the peak scattering
amplitude can also be enhanced or reduced.  The enhancement is
less than the previous prediction. Furthermore, wave propagation
through a plane bubble array can be significantly inhibited in a
range of frequencies slightly higher than the natural frequency of
a single bubble. The results from this paper can also be extended
to scattering by multiple plane arrays of bubbles.

\section*{Acknowledgment}

The work received support from the National Science Council of
Republic of China. The useful comments from referees are greatly
appreciated.


\newpage
\subsection*{Figure Captions}

\begin{description}
\item[Fig. 1] The geometry for a line array of bubbles. The bubble
radius is $a$ and the separation between bubbles is $d$. The
incident wave makes an angle $\theta$ with the line array.

\item[Fig. 2] The geometry for a plane array. The array is in the $x -
y$ plane and forms a square lattice with constant $d$. The
incident wave makes angles $\theta$ and $\phi$ in the spherical
coordinates. In the numerical computation, we consider the case
that the wave propagates along the $z$-axis.

\item[Fig. 3] Line array: Scattering and effective scattering strength with
respect to a single free bubble versus frequency for various
bubble separations.  The bubble radius is 1 mm. Here the notation
$f$ can either refer to the scattering function of a single bubble
or the effective scattering function $F$ of a single bubble
embedded in a line array of bubbles.

\item[Fig. 4] Line array: Scattering strength as a function of $ka$ for different
bubble separations and incidence angles. The bubble
radius is 1 mm.

\item[Fig. 5] Line array: Relative frequency and damping shifts as a function of
$d/a$ for two bubble sizes and two incident angles ($\theta =
\pi/2$ and $\pi/4$). Here $\delta f$ is shift in peak position due
to multiple scattering between bubbles, $f_0$ is the resonance
frequency of a single bubble in isolation ($f_0 = \omega_0/2\pi$).

\item[Fig. 6] Line array: Relative peak scattering amplitude as a function of
$d/a$ for two bubble sizes and two incident angles.

\item[Fig. 7] Plane array: Scattering strength versus frequency for
various bubble separations. The bubble radius is 1 mm. Here the
notation $f$ can either refer to the scattering function of a
single bubble or the effective scattering function $F$ of a single
bubble embedded in a line array of bubbles.

\item[Fig. 8] Plane array: Relative frequency and damping shifts as a function of
$d/a$ for two bubble sizes.

\item[Fig. 9] Plane array: Relative peak scattering amplitude as a function of
$d/a$ for two bubble sizes.

\item[Fig. 10] Plane array: Relative transmission versus $ka$ for four bubble separations.
Without the plane array, the transmission is normalized as one.

\end{description}

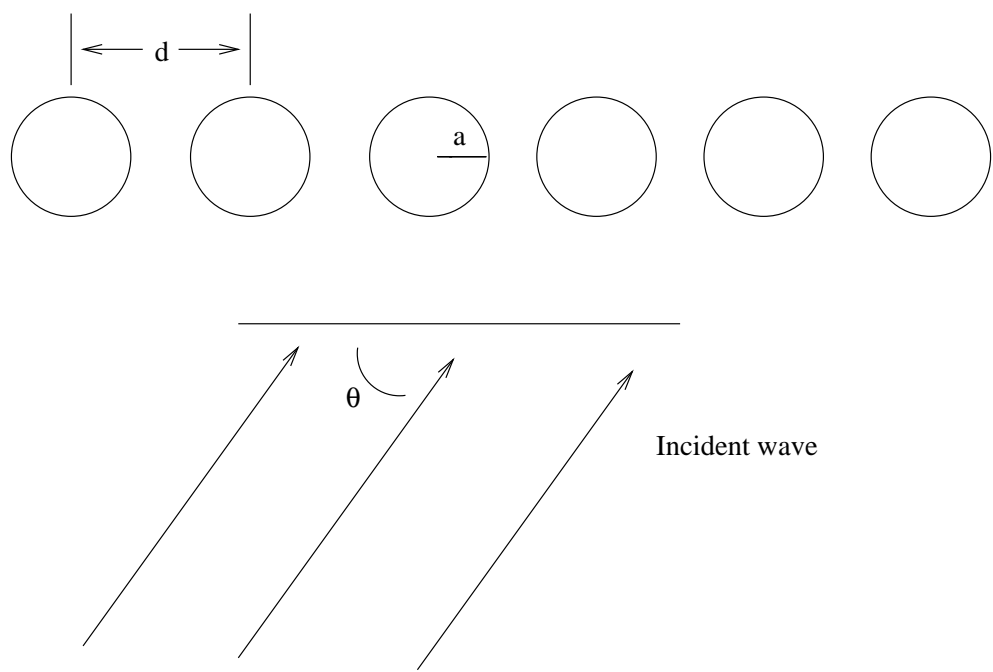
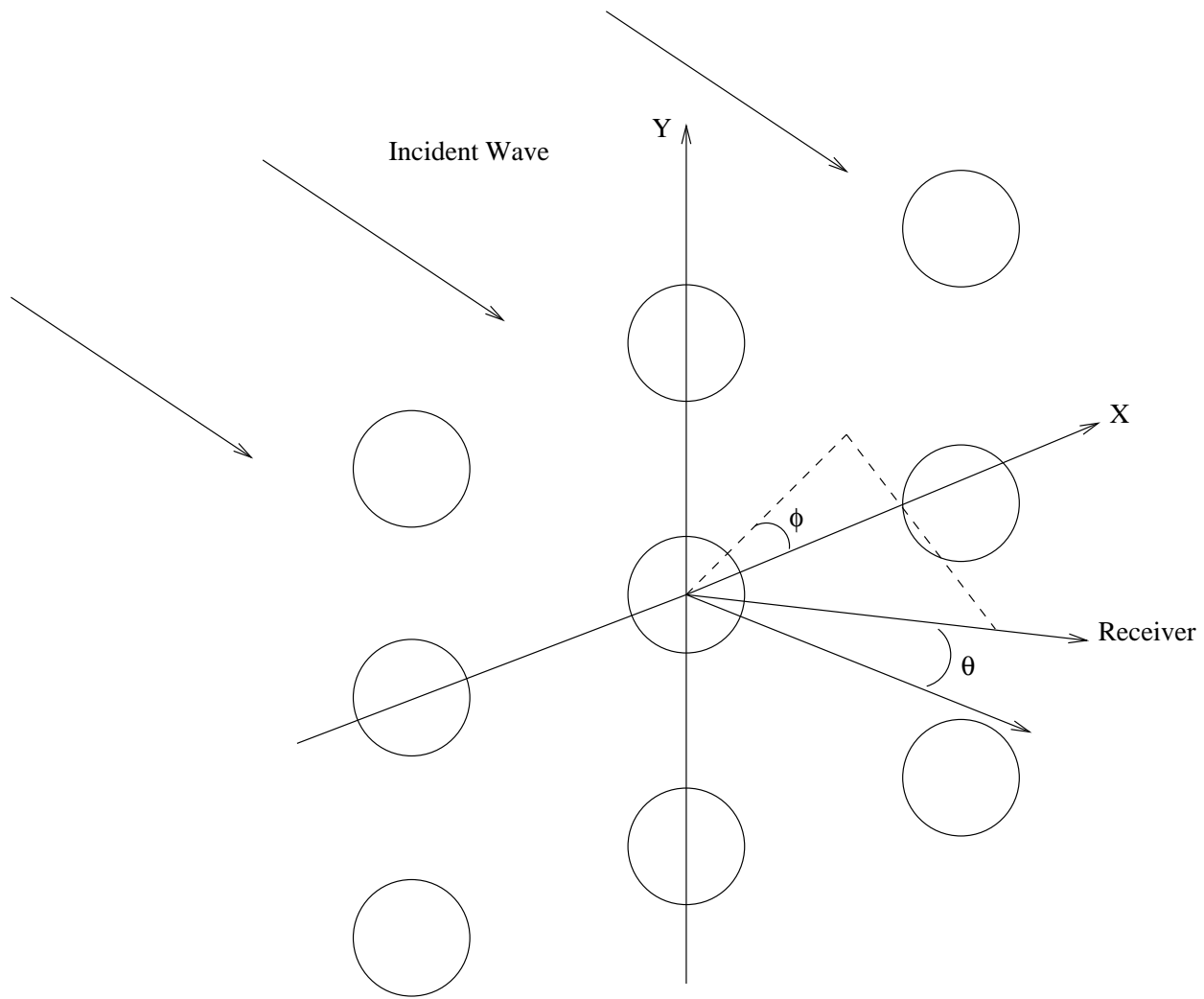
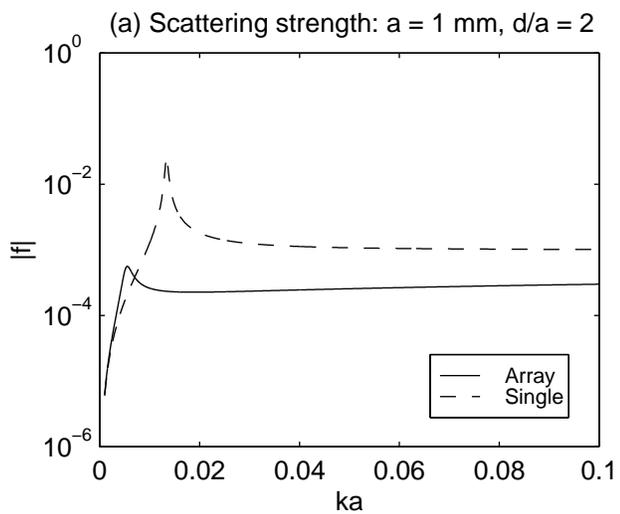
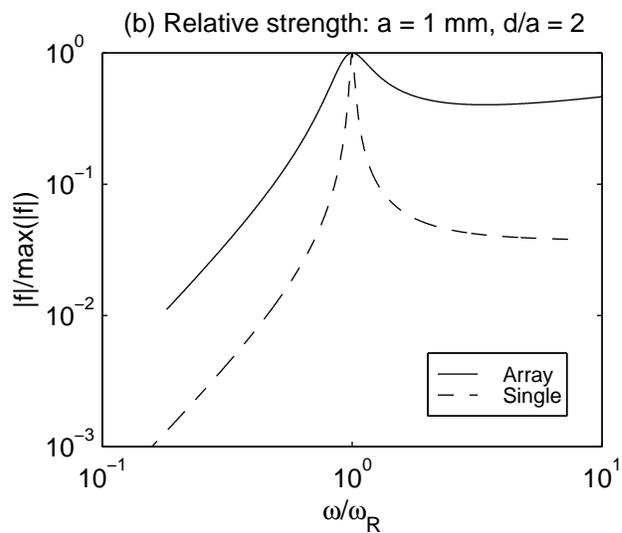
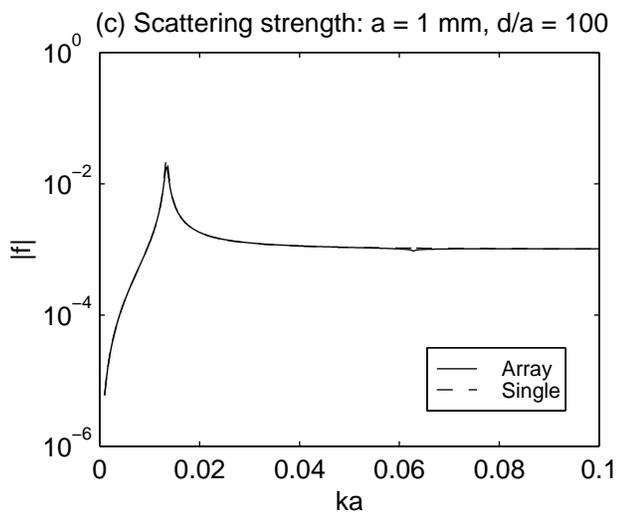
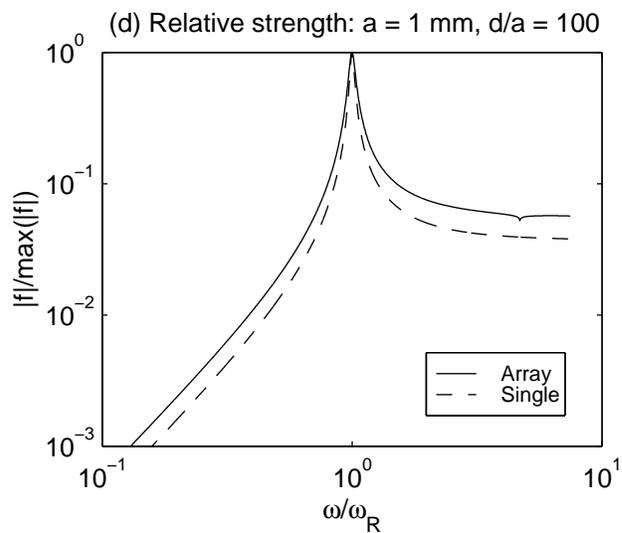
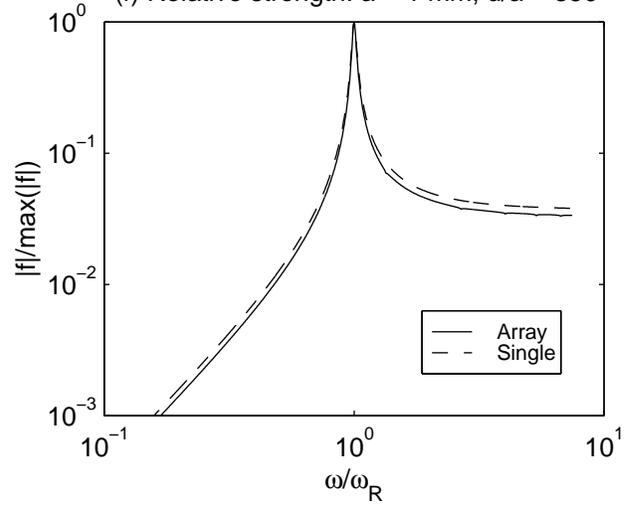
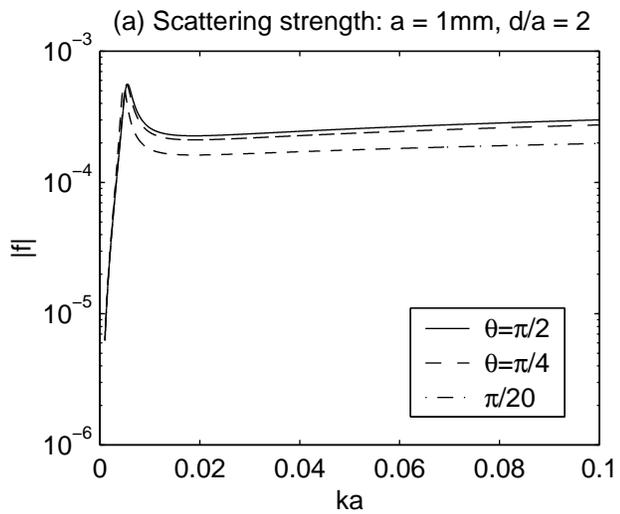
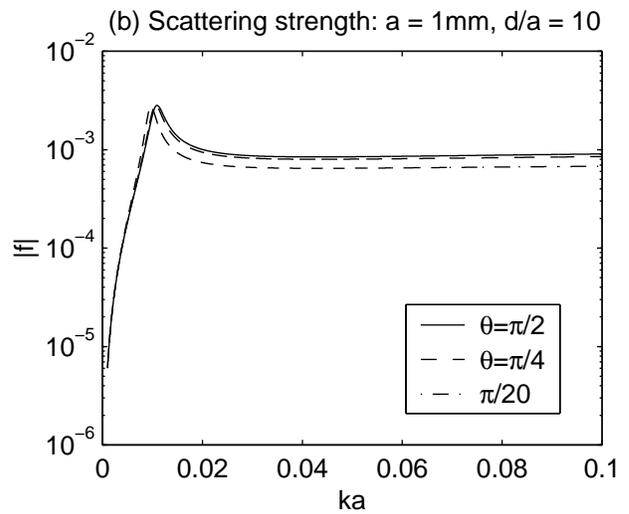
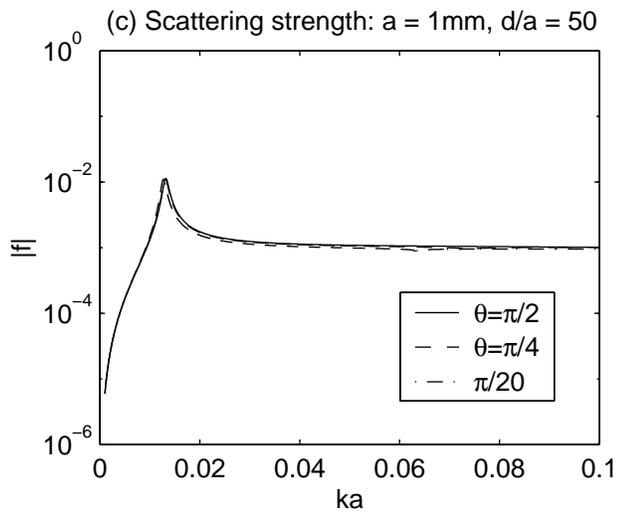
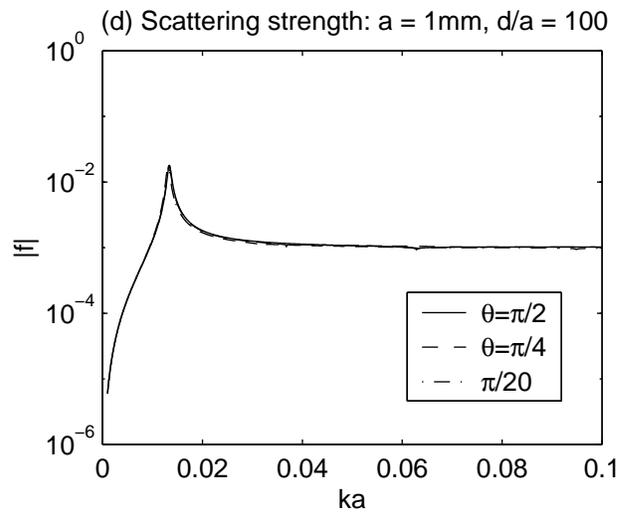
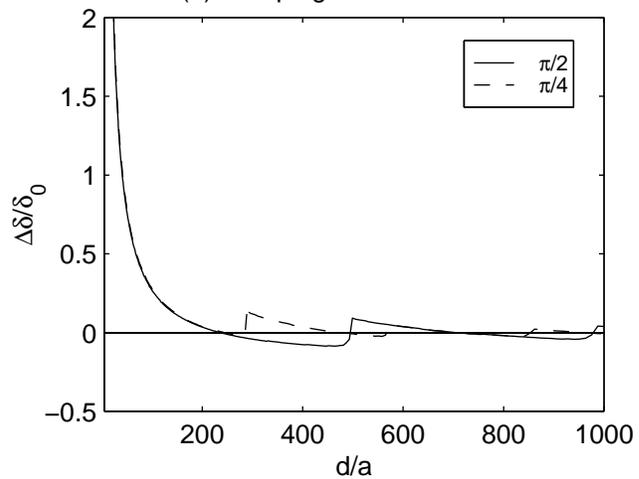
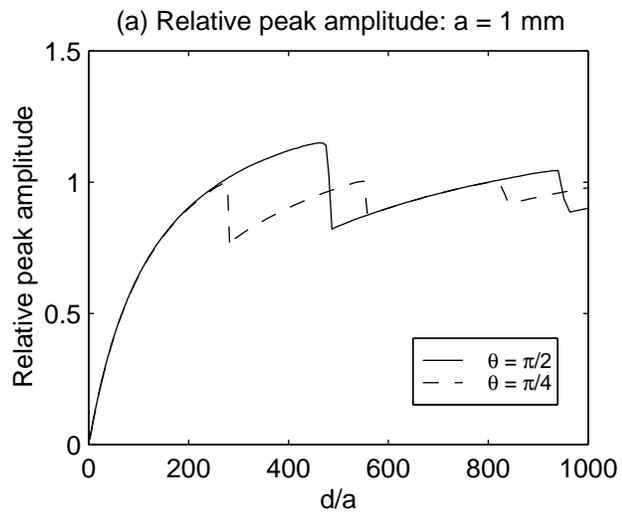
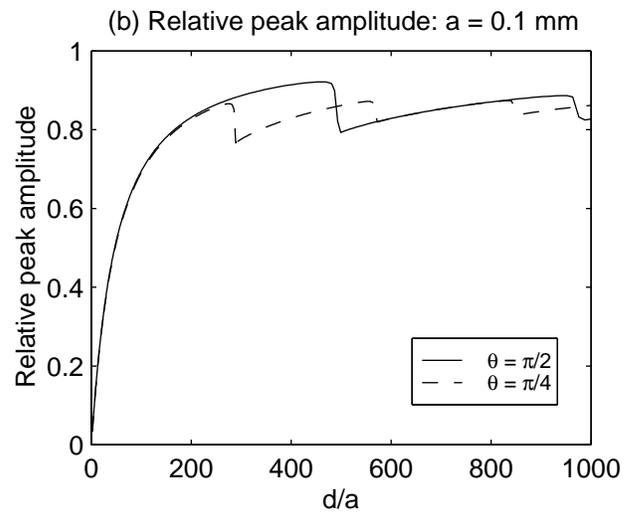
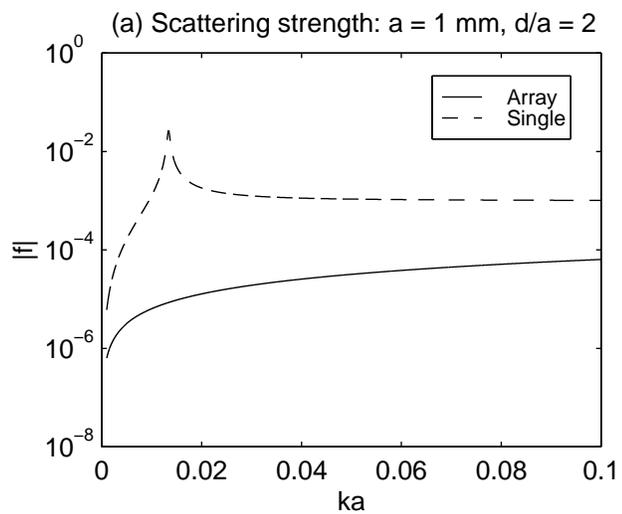
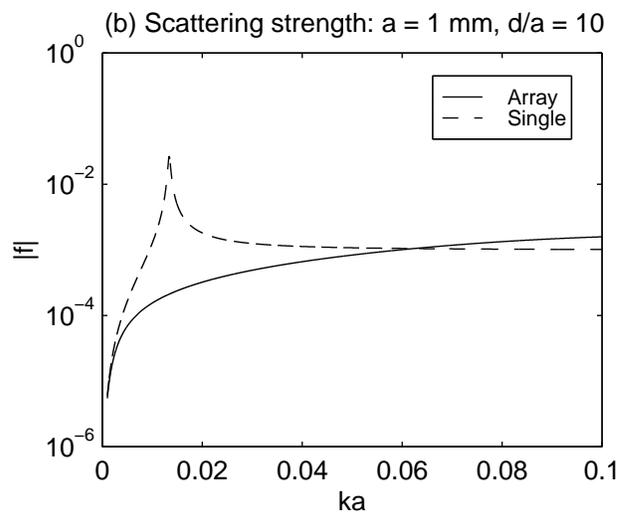
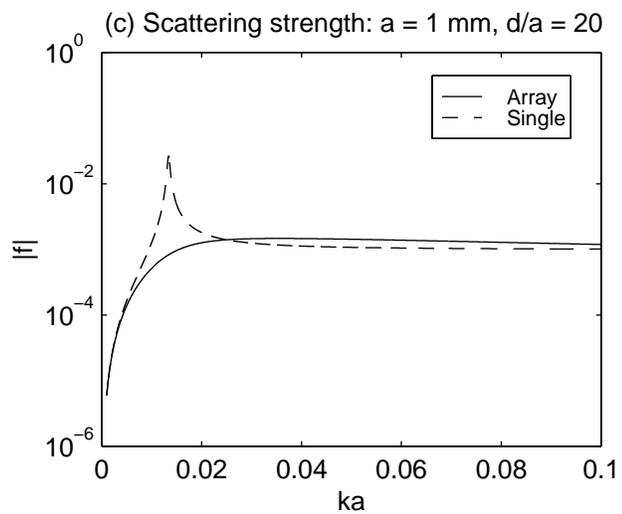
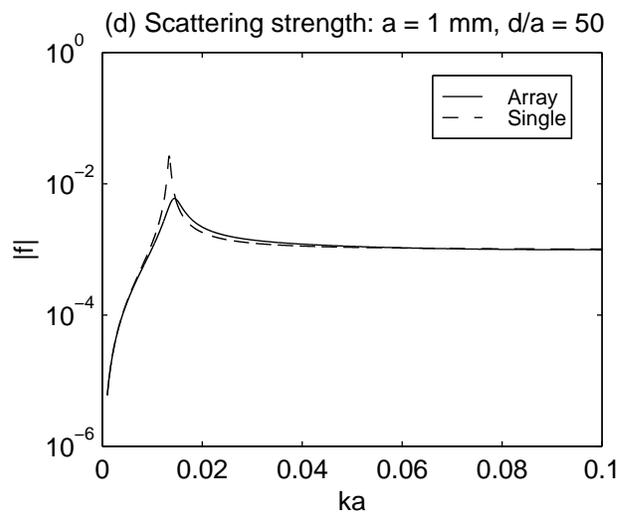
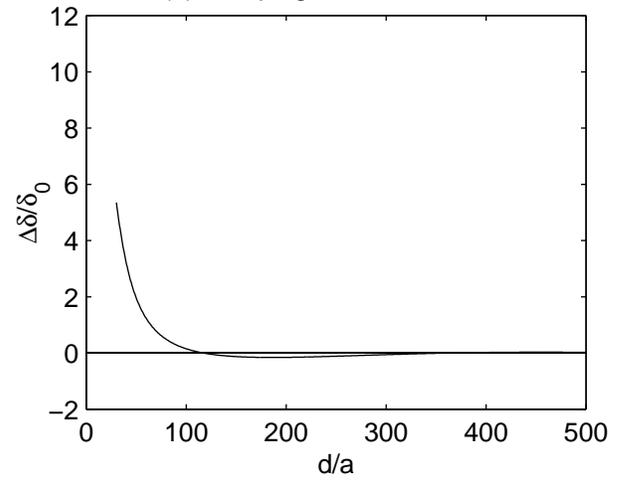
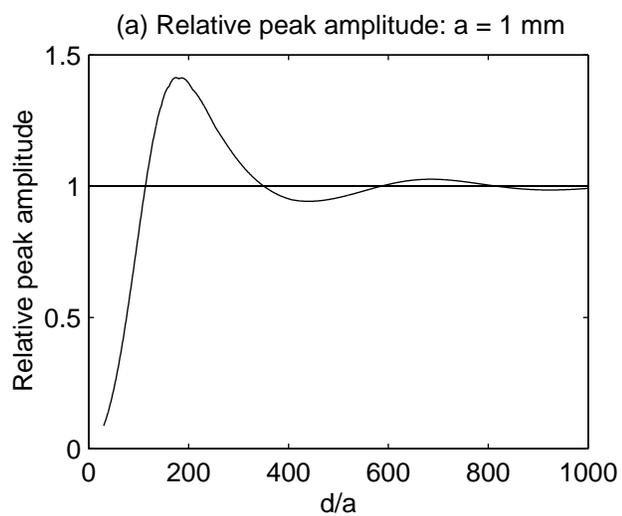
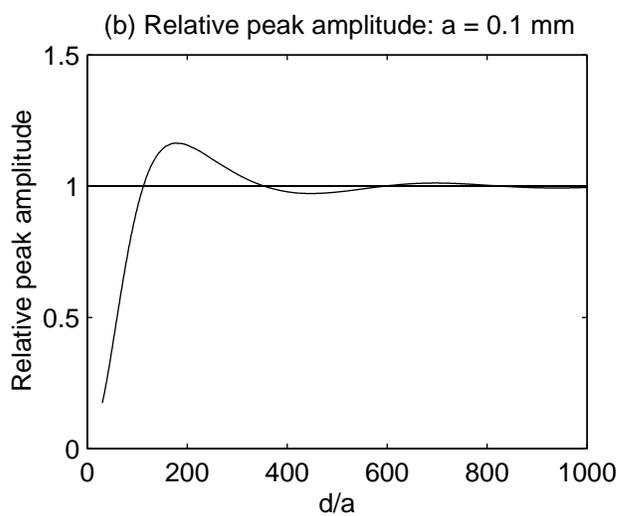
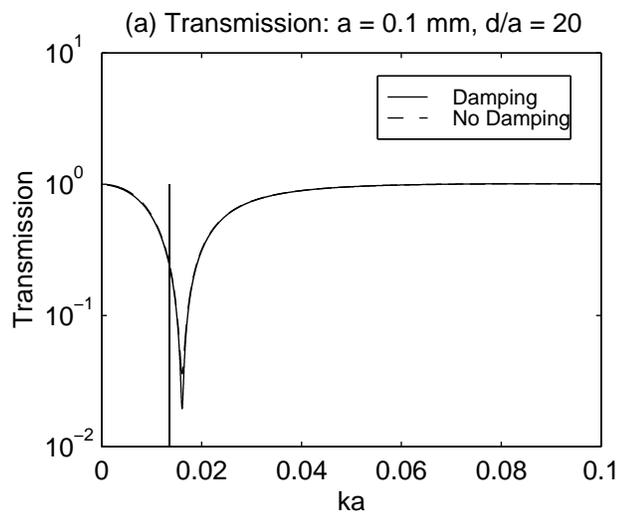
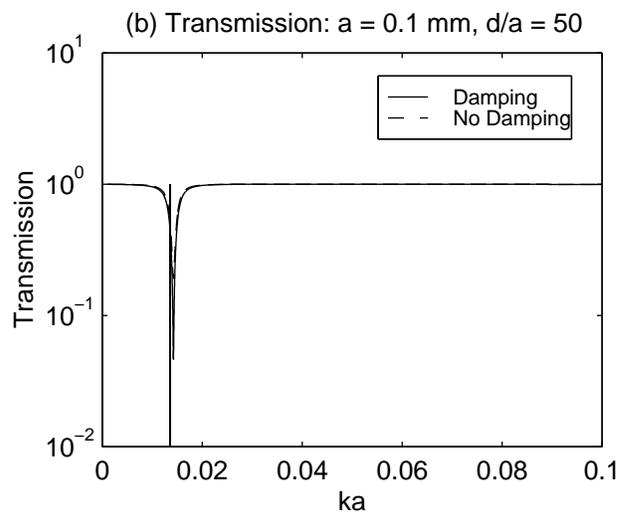
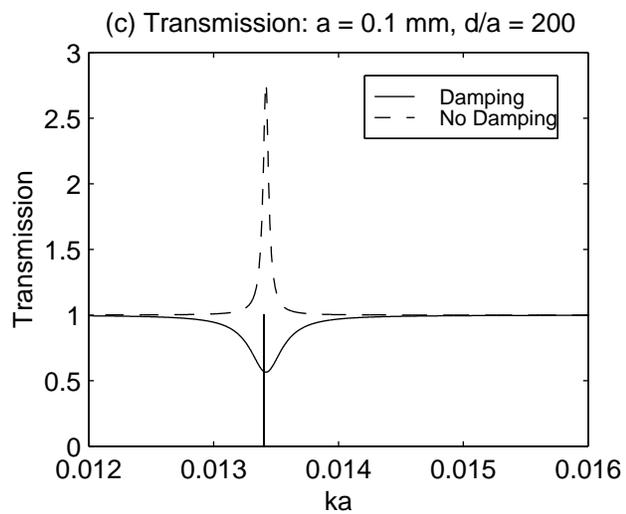
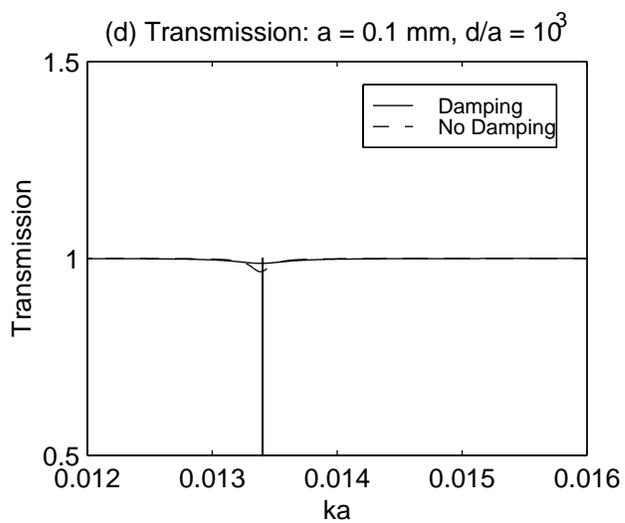
\end{document}